\documentclass[12pt]{article}
\usepackage{amsmath,amsfonts,amssymb}

\begin{document}
\thispagestyle{empty}

\def\theequation{\arabic{section}.\arabic{equation}}
\def\a{\alpha}
\def\b{\beta}
\def\g{\gamma}
\def\d{\delta}
\def\dd{\rm d}
\def\e{\epsilon}
\def\ve{\varepsilon}
\def\z{\zeta}

\def\B{\mbox{\bf B}}

\begin{titlepage}

\renewcommand{\thefootnote}{\fnsymbol{footnote}}
\begin{center}
{\Huge Finite-size Effects for Single Spike}
\end{center}
\vskip 1.2cm \centerline{\Large Changrim  Ahn and P. Bozhilov
\footnote{On leave from Institute for Nuclear Research and Nuclear
Energy, Bulgarian Academy of Sciences, Bulgaria.}}

\vskip 10mm


\centerline{\sl Department of Physics}
\centerline{\sl Ewha Womans University}
\centerline{\sl DaeHyun 11-1, Seoul 120-750, S. Korea} \vspace*{0.6cm}
\centerline{\tt ahn@ewha.ac.kr, bozhilov@inrne.bas.bg}

\vskip 20mm

\baselineskip 18pt

\begin{center}
{\bf Abstract}
\end{center}
We use the reduction of the string dynamics on $R_t\times S^3$ to
the Neumann-Rosochatius integrable system to map {\it
all} string solutions described by this dynamical system onto
solutions of the complex sine-Gordon integrable model. This
mapping relates the parameters in the solutions on both sides of
the correspondence. In the framework of this approach, we find
finite-size string solutions, their images in the (complex)
sine-Gordon system, and the leading finite-size effects of
the single spike ``$E-\Delta\varphi$'' relation for both
$R_t\times S^2$ and $R_t\times S^3$ cases.

\end{titlepage}
\newpage
\baselineskip 18pt

\def\nn{\nonumber}
\def\tr{{\rm tr}\,}
\def\p{\partial}
\newcommand{\bea}{\begin{eqnarray}}
\newcommand{\eea}{\end{eqnarray}}
\newcommand{\bde}{{\bf e}}
\renewcommand{\thefootnote}{\fnsymbol{footnote}}
\newcommand{\be}{\begin{equation}}
\newcommand{\ee}{\end{equation}}
\newcommand{\h}{\hspace{0.5cm}}

\vskip 0cm

\renewcommand{\thefootnote}{\arabic{footnote}}
\setcounter{footnote}{0}

\setcounter{equation}{0}
\section{Introduction}

Recent developments in AdS/CFT correspondence between type IIB
strings on $AdS_5\times S^5$ and its dual ${\cal N}=4$ super
Yang-Mills (SYM) theories \cite{AdS/CFT} are mainly based on the
integrabilities discovered in both theories. Integrability of the
SYM side appears in the calculations of conformal dimensions which
are related to the string energies according to the AdS/CFT
correspondence. A remarkable observation by Minahan and Zarembo
\cite{MZ} is that the conformal dimension of an operator composed
of scalar fields in the ${\cal N}=4$ SYM in the planar limit can
be computed by diagonalizing the Hamiltonian of one-dimensional
integrable spin chain model. This task can be done by solving a
set of coupled algebraic equations called Bethe ansatz equations.
It has been shown that explicit calculation of the eigenvalues for
various SYM operators agree with those computed from the SYM
perturbation theory. This result has been further extended to the
full $PSU(2,2|4)$ sector \cite{Beis} and the Bethe ansatz
equations which are supposed to hold for all loops are conjectured
\cite{BDS,BeiStai}.

The string side of the correspondence is mostly studied at the classical level due to the lack
of full quantization.
The type IIB string theory on $AdS_5\times S^5$ is described by a nonlinear sigma model with
$PSU(2,2|4)$ symmetry \cite{MetTse}.
This sigma model has been shown to have an infinite number of local and nonlocal conserved currents
\cite{Bena} and some of the conserved charges such as energy and angular momentum are
computed explicitly from the classical integrability (see, for example, \cite{Tseytlin}
and the references therein).
These results based on the classical integrability provide valuable information on the AdS/CFT
duality in the domain of large t'Hooft coupling constant.
A new direction to quantizing the string theory is to find exact $S$-matrix between the fundamental
spectrum of the theory on the world sheet.
It has been shown that the $S$-matrix along with the exact particle spectrum can be determined by
the underlying symmetry and integrability in the theory \cite{Beisert,AFZ}.
The overall scalar factor of the $S$-matrix, sometimes called as dressing phase, which can not be
determined by the symmetry alone, has been computed exactly \cite{BHL,BES} from the
crossing relation \cite{Janik}.
In this process, the explicit expression of the dressing phase in the classical limit, which was
determined by the classical integrability, was essential \cite{AFS}.

Various classical solutions play an important role in testing and
understanding the correspondence. The classical giant magnon (GM)
state \cite{HM06} discovered in $R_t\times S^2$ gives a strong
support for the conjectured all-loop $SU(2)$ spin chain and makes
it possible to get a deep insight in the AdS/CFT duality. In
addition, this solution is related to classical sine-Gordon (SG)
model which provides a geometric understanding of the string in
curved space. This is extended to the magnon bound state which
corresponds to a string moving on $R_t \times S^3$ and related to
the complex sine-Gordon (CSG) model \cite{Dorey,CDO06}. Further
extensions to $R_t \times S^5$ have been also worked out
\cite{KRT06,SprVol,Ryangi,DimRas}. Another interesting classical
string solution is the spiky string which has been first found in
the $AdS$ space \cite{Krucz} and in the $S^5$ \cite{Ryangii}. A
particular case of these states is the single spike (SS) which
describes a string which is wrapping infinitely around the equator
with a spike in the middle. This has been investigated in a static
gauge using the Nambu-Goto action in $S^2$ and $S^3$ by Ishizeki
and Kruczenski \cite{IK07}. In addition, they have shown that both
the GM and the SS solutions on $S^2$ can be related to the
classical SG equation. This is possible by reformulating
the problem in a conformal gauge using the Polyakov action and
assuming a particular ansatz for string coordinates which leads to
the well-known one-dimensional integrable Neumann-Rosochatius (NR)
system. This approach has been previously developed and applied to
find classical solutions in \cite{AFRT,ART,KRT06}. The application
of the NR system to the SS in $S^3$ has been worked out in
\cite{BobR07}.
Also a semiclassical quantization of the SS has been recently studied in \cite{AbbAni}

More recently the finite-size correction, or the L\"uscher correction, is actively investigated as a
new window for the AdS/CFT correspondence.
The integrable spin chains based on the Bethe ansatz are showing several limitations such as
the wrapping problem which occurs in dealing with a composite operator of a finite length in
a strong t'Hooft coupling limit.
It is quite important to compute the conformal dimensions of such operators and compare with
the energies of the string states with all other quantum numbers finite.
One way of confirming the $S$-matrix is to derive the L\"uscher correction from the $S$-matrix and
compare with the classical string result.
The finite-size effects for the GM have been computed from the $S$-matrix \cite{Janikii}
and are shown to be consistent with classical string results.
Finite-size effects for the GM has been first found by solving the string sigma model in a uniform and conformal
gauges \cite{AFZ06} and, subsequently, many related results, such as gauge independence \cite{AFGS},
multi GM states \cite{MinSax} and quantization of finite-size GM \cite{RamSem}, have been derived.
This result has been also related to explicit solutions of the SG equation
in a finite-size space \cite{KM0803}.

In this article we compute the finite-size effects for the SS in
both $S^2$ and $S^3$.
The finite-size SS solutions were related to the helical string solutions
without analyzing the corrections to the energy-charge dispersion relations  in \cite{Okamura}.
While there are some conjectures for the
quantum SS from the integrable spin chain models such as the
Hubbard model \cite{IK07} or antiferromagnetic $SO(6)$ spin chain
model \cite{Okamura}, the $S$-matrix for the SS is known only in
the classical limit \cite{IKSV07} and still not available at the
full quantum level. This excludes the $S$-matrix approach for the
L\"uscher correction and leaves the classical analysis as a viable
option.

The paper is organized as follows. In sect.2 we introduce the
classical string action on $R_t\times R^3$ and the corresponding
NR system. This system is shown to be equivalent to a particular
case of the CSG equation in sect.3.
In sect.4 we provide our main result on the finite-size effects of
the SS on both $R_t\times S^2$ and $R_t\times S^3$.
We conclude the paper with some remarks
in sect.5 and with Appendices containing the explicit relationship
between the NR system and CSG for the GM and SS in infinite-size
system and $\epsilon$-expansions for elliptic functions and relevant coefficients.

\setcounter{equation}{0}
\section{Strings on $R_t\times S^3$ and the NR Integrable System}

Let us start with the Polyakov string action \bea &&S^P=
-\frac{T}{2}\int d^2\xi\sqrt{-\gamma}\gamma^{ab}G_{ab},\h G_{ab} =
g_{MN}\p_a X^M\p_bX^N,\\ \nn &&\p_a=\p/\p\xi^a,\h a,b = (0,1),
\h(\xi^0,\xi^1)=(\tau,\sigma),\h M,N = (0,1,\ldots,9),\eea and
choose {\it conformal gauge} $\gamma^{ab}=\eta^{ab}=diag(-1,1)$,
in which the Lagrangian and the Virasoro constraints take the form
\bea\label{l}
&&\mathcal{L}_s=\frac{T}{2}\left(G_{00}-G_{11}\right) \\
\label{00} && G_{00}+G_{11}=0,\qquad G_{01}=0.\eea where
$T$ is the string tension.

We embed the string in $R_t\times S^3$ subspace of $AdS_5\times
S^5$ as follows \bea\nn Z_0=Re^{it(\tau,\sigma)},\h
W_j=Rr_j(\tau,\sigma)e^{i\varphi_j(\tau,\sigma)},\h
\sum_{j=1}^{2}W_j\bar{W}_j=R^2,\eea where $R$ is the common radius
of $AdS_5$ and $S^5$, and $t$ is the $AdS$ time. For this
embedding, the metric induced on the string worldsheet is given by
\bea\nn G_{ab}=-\p_{(a}Z_0\p_{b)}\bar{Z}_0
+\sum_{j=1}^{2}\p_{(a}W_j\p_{b)}\bar{W}_j=R^2\left[-\p_at\p_bt +
\sum_{j=1}^{2}\left(\p_ar_j\p_br_j +
r_j^2\p_a\varphi_j\p_b\varphi_j\right)\right].\eea
The corresponding string Lagrangian becomes \bea\nn
\mathcal{L}=\mathcal{L}_s +
\Lambda_s\left(\sum_{j=1}^{2}r_j^2-1\right),\eea where $\Lambda_s$
is a Lagrange multiplier.
In the case at hand, the background metric does not depend on $t$
and $\varphi_j$. Therefore, the conserved quantities
are the string energy $E_s$ and two angular momenta $J_j$, given
by \bea\label{gcqs} E_s=-\int
d\sigma\frac{\p\mathcal{L}_s}{\p(\p_0 t)},\h J_j=\int
d\sigma\frac{\p\mathcal{L}_s}{\p(\p_0\varphi_j)}.\eea

It is known that restricting ourselves to the case
\bea\label{NRA} &&t(\tau,\sigma)=\kappa\tau,\h
r_j(\tau,\sigma)=r_j(\xi),\h
\varphi_j(\tau,\sigma)=\omega_j\tau+f_j(\xi),\\ \nn
&&\xi=\alpha\sigma+\beta\tau,\h \kappa, \omega_j, \alpha,
\beta=constants,\eea
reduces the problem to solving the NR integrable system \cite{KRT06}.
For the case under consideration, the NR Lagrangian reads (prime
is used for $d/d\xi$) \bea\label{LNR} L_{NR}=(\alpha^2-\beta^2)
\sum_{j=1}^{2}\left[r_j'^2-\frac{1}{(\alpha^2-\beta^2)^2}
\left(\frac{C_j^2}{r_j^2} + \alpha^2\omega_j^2r_j^2\right)\right]
+\Lambda_s\left(\sum_{j=1}^{2}r_j^2-1\right),\eea where the
parameters $C_j$ are integration constants after single time
integration of the equations of motion for $f_j(\xi)$:
\bea
f_j'=\frac{1}{\alpha^2-\beta^2}\left(\frac{C_j}{r_j^2} +
\beta\omega_j\right).\label{fjprime}
\eea
The constraints (\ref{00})
give the conserved Hamiltonian $H_{NR}$ and a relation between the
embedding parameters and the arbitrary constants $C_j$:
\bea\label{HNR} &&H_{NR}=(\alpha^2-\beta^2)
\sum_{j=1}^{2}\left[r_j'^2+\frac{1}{(\alpha^2-\beta^2)^2}
\left(\frac{C_j^2}{r_j^2} + \alpha^2\omega_j^2r_j^2\right)\right]
=\frac{\alpha^2+\beta^2}{\alpha^2-\beta^2}\kappa^2,
\\ \label{01R} &&\sum_{j=1}^{2}C_j\omega_j + \beta\kappa^2=0.\eea
For closed strings, $r_j$ and $f_j$
satisfy the following periodicity conditions
\bea
r_j(\xi+2\pi\alpha)=r_j(\xi),\h f_j(\xi+2\pi\alpha)=f_j(\xi)+2\pi n_\alpha,\label{pbc}
\eea
where $n_\alpha$ are integer winding numbers.
On the ansatz (\ref{NRA}), $E_s$ and $J_j$ introduced in
(\ref{gcqs}) take the form \bea\label{cqs} E_s=
\frac{\sqrt{\lambda}}{2\pi}\frac{\kappa}{\alpha}\int d\xi,\h J_j=
\frac{\sqrt{\lambda}}{2\pi}\frac{1}{\alpha^2-\beta^2}\int d\xi
\left(\frac{\beta}{\alpha}C_j+\alpha\omega_j r_j^2\right),\eea
where we have used that the string tension and the 't Hooft
coupling constant $\lambda$ are related by
$TR^2=\frac{\sqrt{\lambda}}{2\pi}$.

In order to identically satisfy the embedding condition \bea\nn
\sum_{j=1}^{2}r_j^2-1=0,\eea we introduce a new variable $\theta(\xi)$ by
\bea r_1(\xi)=\sin{\theta(\xi)},\h r_2(\xi)=\cos{\theta(\xi)}.\label{sincos}\eea
Then, Eq.(\ref{HNR}) leads to
\bea\label{tsol}
\theta'(\xi)&=&\pm\frac{1}{\alpha^2-\beta^2}
\left[(\alpha^2+\beta^2)\kappa^2 - \frac{C_1^2}{\sin^2{\theta}} -
\frac{C_2^2}{\cos^2{\theta}} -
\alpha^2\left(\omega_1^2\sin^2{\theta}
+\omega_2^2\cos^2{\theta}\right)\right]^{1/2}\\
&\equiv& \pm\frac{1}{\alpha^2-\beta^2}\ \Theta(\theta),\nonumber
\eea
which can be integrated to give
\bea
\xi(\theta)= \pm(\alpha^2-\beta^2)\int\frac{d\theta}{\Theta(\theta)}.
\eea
From Eqs.(\ref{fjprime}) and (\ref{sincos}), we can obtain
\bea\label{f1s}
&&f_1=\frac{\beta\omega_1\xi}{\alpha^2-\beta^2}\pm
C_1\int\frac{d\theta}{\sin^2\theta\ \Theta(\theta)},\\ \label{f2s}
&&f_2=\frac{\beta\omega_2\xi}{\alpha^2-\beta^2}\pm
C_2\int\frac{d\theta}{\cos^2\theta\ \Theta(\theta)}.
\eea
Let us also point out that
the solutions for $\xi(\theta)$ and $f_j$ must satisfy the
conditions (\ref{01R}) and (\ref{pbc}).
All these solve formally the NR system for the present case.

As far as we are searching for real solutions, the expressions
under the square roots in (\ref{tsol}) must be positive, which put restrictions on the
possible values of the parameters. Of course, this condition
arises from the requirement that the NR Hamiltonian (\ref{HNR}) should be
positive.

\setcounter{equation}{0}
\section{Relationship between the NR and CSG Systems}

Due to Pohlmeyer \cite{P76}, we know that the string dynamics on
$R_t\times S^3$ can be described by the CSG
equation. In this section, we derive the relation between
the solutions of the two integrable systems.

The CSG system is defined by the Lagrangian
\bea\nn
\mathcal{L}(\psi) =
\frac{\eta^{ab}\p_a\bar{\psi}\p_b\psi}{1-\bar{\psi}\psi} +
M^2\bar{\psi}\psi
\eea
which give the equation of motion
\bea\nn \p_a\p^a\psi
+\bar{\psi}\frac{\p_a\psi\p^a\psi}{1-\bar{\psi}\psi} -
M^2(1-\bar{\psi}\psi)\psi=0.\eea
If we represent $\psi$ in the
form \bea\nn \psi=\sin(\phi/2)\exp(i\chi/2),\eea the Lagrangian
can be expressed as \bea\nn
\mathcal{L}(\phi,\chi)=\frac{1}{4}\left[\p_a\phi\p^a\phi +
\tan^2(\phi/2)\p_a\chi\p^a\chi + (2M)^2\sin^2(\phi/2)\right],\eea
along with the equations of motion
\bea\label{fem} &&\p_a\p^a\phi -
\frac{1}{2}\frac{\sin(\phi/2)}{\cos^3(\phi/2)}\p_a\chi\p^a\chi -
M^2\sin\phi=0,\\ \label{kem} &&\p_a\p^a\chi +
\frac{2}{\sin\phi}\p_a\phi\p^a\chi=0.\eea The SG system
corresponds to a particular case of $\chi=0$.

To relate the NR system with the CSG integrable system, we consider the case
\bea\nn \phi=\phi(\xi),\h \chi=A\sigma+B\tau + \tilde{\chi}(\xi),
\eea
where $\phi$ and $\tilde{\chi}$ depend on
only one variable $\xi=\alpha\sigma+\beta\tau$ in the same way as in our NR ansatz (\ref{NRA}).
Then the equations of
motion (\ref{fem}), (\ref{kem}) reduce to
\bea\label{fer} &&\phi''
- \frac{1}{2}\frac{\sin(\phi/2)}{\cos^3(\phi/2)}
\left[\tilde{\chi}'^2 +
2\frac{A\alpha-B\beta}{\alpha^2-\beta^2}\tilde{\chi}' +
\frac{A^2-B^2}{\alpha^2-\beta^2}\right]
- \frac{M^2\sin\phi}{\alpha^2-\beta^2}=0,\\
\label{ker} &&\tilde{\chi}'' +
\frac{2\phi'}{\sin\phi}\left(\tilde{\chi}' +
\frac{A\alpha-B\beta}{\alpha^2-\beta^2}\right)=0.\eea

We further restrict ourselves to the case of $A\alpha=B\beta$.
A trivial solution of Eq.(\ref{ker}) is $\tilde{\chi}=constant$,
which corresponds to the solutions of the CSG equations considered in \cite{CDO06,OS06}
for a GM string on $R_t\times S^3$.
More nontrivial solution of (\ref{ker}) is
\bea\label{kfi}
\tilde{\chi}= C_\chi\int \frac{d\xi}{\tan^2(\phi/2)}.\eea
The replacement of the above into (\ref{fer}) gives
\bea\label{fef}\phi''=\frac{M^2\sin\phi}{\alpha^2-\beta^2} +
\frac{1}{2}\left[C_\chi^2\frac{\cos(\phi/2)}{\sin^3(\phi/2)}
-\frac{A^2}{\beta^2}\frac{\sin(\phi/2)}{\cos^3(\phi/2)}\right].\eea
Integrating once, we obtain
\bea\label{ffi} \phi'&=&\pm\left[\left(C_\phi -
\frac{2M^2}{\alpha^2-\beta^2}\right) +
\frac{4M^2}{\alpha^2-\beta^2}\sin^2(\phi/2)
-\frac{A^2/\beta^2}{1-\sin^2(\phi/2)} -
\frac{C_\chi^2}{\sin^2(\phi/2)}\right]^{1/2}\\
&\equiv&\pm\Phi(\phi),\nonumber
\eea
from which we get
\bea\nn \xi(\phi)=\pm\int \frac{d\phi}{\Phi(\phi)},\qquad\chi(\phi)=
\frac{A}{\beta}\left(\beta\sigma+\alpha\tau\right)\pm C_\chi\int
\frac{d\phi}{\tan^2(\phi/2)\Phi(\phi)}.\eea
All these solve
the CSG system for the considered particular case.
It is clear from (\ref{ffi}) that the expression inside the square root must be positive.

Now we are ready to establish a correspondence between the NR and
CSG integrable systems described above. To this end, we make the
following identification \bea\label{equiv} \sin^2(\phi/2)\equiv
\frac{\sqrt{-G}}{K^2}\eea where $G$ is the determinant of the
induced metric $G_{ab}$ computed on the constraints (\ref{00})
and $K^2$ is a parameter which will be fixed
later\footnote{For $K^2=\kappa^2$, this definition of the angle
$\phi$ coincides with the one used in \cite{CDO06}, which is based
on the Pohlmeyer's reduction procedure \cite{P76}.}. For our NR
system, $\sqrt{-G}$ is given by \bea\label{equiv1} \sqrt{-G}
=\frac{R^2\alpha^2}{\alpha^2-\beta^2}\left[(\kappa^2-\omega_1^2) +
(\omega_1^2-\omega_2^2)\cos^2\theta\right].\eea

We want the field $\phi$, defined in (\ref{equiv}) through NR
quantities, to {\it identically} satisfy (\ref{ffi}) derived from
the CSG equations. This imposes relations between the parameters
involved, which are given in appendix A. In this way, we mapped
{\it all} string solutions on $R_t\times S^3$ (in particular on
$R_t\times S^2$) described by the NR integrable system onto
solutions of the CSG (in particular SG) equations.
From (\ref{equiv2}) one can see that the parameters $A$ and $C_\chi$
are nonzero in general on $R_t\times S^2$ where $\omega_2=C_2=0$.
This means that there exist string
solutions on $R_t\times S^2$ which correspond to solutions of
the CSG system. Only when $M^2=\kappa^2$, all string solutions on
$R_t\times S^2$ are represented by solutions of the SG equation.

For the GM and SS solutions, which we are interested
in, the relations between the NR and CSG parameters simplify a
lot. Let us write them explicitly.
The GM solutions correspond to $C_2=0$,
$\kappa^2=\omega_1^2$. This leads to \bea\label{GMp}
&&C_\phi= \frac{2}{\alpha^2-\beta^2}\left[3M^2-
2\left(\omega_1^2
-\frac{\omega_2^2}{1-\beta^2/\alpha^2}\right)\right],\h
K^2=R^2M^2,
\\ \nn &&A^2=\frac{4}{\alpha^2/\beta^2-1}\left(M^2-\omega_1^2
+ \frac{\omega_2^2}{1-\beta^2/\alpha^2}\right), \h
C_{\chi}=0.\eea Therefore, for all GM strings the field
$\chi$ is linear function at most. Since $A=C_{\chi}=0$
implies $\chi=0$, it follows from here that there exist GM
string solutions on $R_t\times S^3$, which are mapped not on CSG
solutions but on SG solutions instead. This happens exactly when
\bea\nn M^2=\omega_1^2 -
\frac{\omega_2^2}{1-\beta^2/\alpha^2}.\eea In that case the
nonzero parameters are \bea\nn K^2=R^2\left(\omega_1^2 -
\frac{\omega_2^2}{1-\beta^2/\alpha^2}\right),\h C_\phi=
\frac{2}{\alpha^2-\beta^2}\left(\omega_1^2 -
\frac{\omega_2^2}{1-\beta^2/\alpha^2}\right),\eea and the
corresponding solution of the SG equation  can be found from
(\ref{ffi}) to be \bea\label{S3SG} \sin(\phi/2)=
\frac{1}{\cosh\left[\sqrt{\frac{\omega_1^2-\omega_2^2/
\left(1-\beta^2/\alpha^2\right)}{1-\beta^2/\alpha^2}}
\left(\sigma+ \frac{\beta}{\alpha}\tau\right)-\eta_0\right]},\h
\eta_0=const.\eea
Replacing (\ref{S3SG}) in (\ref{equiv}),
(\ref{equiv1}), one obtains the GM solution (\ref{gmsol}) as it should be.

For the SS solutions $C_2=0$,
$\kappa^2=\omega_1^2\alpha^2/\beta^2$. This results in \bea\nn
&&C_\phi= \frac{2}{\beta^2-\alpha^2}\left[
2\left(2\omega_1^2\alpha^2/\beta^2
+\frac{\omega_2^2}{\beta^2/\alpha^2-1}\right)-3M^2\right], \\
\label{SSp}
&&A^2=\frac{4}{M^4(1-\alpha^2/\beta^2)}\left(\omega_1^2\alpha^2/\beta^2
-M^2\right)^2
\left(\frac{\omega_2^2}{\beta^2/\alpha^2-1}-M^2\right), \\ \nn
&&C_{\chi}=\frac{2\omega_1^2\omega_2\alpha^3}
{M^2(\beta^2-\alpha^2)\beta^2},\h K^2=R^2M^2.\eea We want to point
out that $C_{\chi}$ is always nonzero on $S^3$ contrary to the GM
case, which makes $\chi$ also non-vanishing. To our knowledge, the
CSG solutions corresponding to the SS on $R_t\times S^3$ are not
given in the literature. To study this problem, we will consider
the case when $A=0$. $A$ can be zero when \bea\label{asszo}
M^2=\kappa^2=\omega_1^2\alpha^2/\beta^2\h \mbox{or}\h
M^2=\frac{\omega_2^2}{\beta^2/\alpha^2-1},\eea As is seen from
(\ref{asszo}), we have two options, and we restrict ourselves to
the first one\footnote{It turns out that the second option does
not allow real solutions.}. Replacing
$M^2=\omega_1^2\alpha^2/\beta^2$ in (\ref{SSp}) and using the
resulting expressions for $C_\phi$ and $C_{\chi}$ in (\ref{ffi}),
one obtains the simplified equation \bea\nn \phi'^2=\frac{4}
{\beta^2-\alpha^2}
\left[\omega_1^2\frac{\alpha^2}{\beta^2}\cos^2(\phi/2) -
\frac{\omega_2^2}{\beta^2/\alpha^2-1}\cot^2(\phi/2)\right]\eea
with solution \bea\label{phis} \sin^2(\phi/2)=
\tanh^2\left(C\xi\right) +
\frac{\omega_2^2}{\omega_1^2\left(1-\alpha^2/\beta^2\right)\cosh^2
\left(C\xi\right)},\eea where \bea\nn
C=\frac{\alpha\omega_1\sqrt{1-\alpha^2/\beta^2-\omega_2^2/\omega_1^2}}
{\beta^2\left(1-\alpha^2/\beta^2\right)}.\eea This agrees with
Eqs. (\ref{equiv}) and (\ref{equiv1}). By inserting (\ref{phis})
into (\ref{kfi}) one can find \bea\nn
\chi=\tilde{\chi}=2\arctan\left[\frac{\omega_1}{\omega_2}
\sqrt{1-\alpha^2/\beta^2-\omega_2^2/\omega_1^2}
\tanh\left(C\xi\right)\right].\eea Hence, the CSG field $\psi$ for
the case at hand is given by \bea\label{CSGsol}
\psi&=&\sqrt{\tanh^2\left(C\xi\right) +
\frac{\omega_2^2}{\omega_1^2\left(1-\alpha^2/\beta^2\right)\cosh^2
\left(C\xi\right)}}\\
\nn &&\times \exp\left\{i\arctan\left[\frac{\omega_1}{\omega_2}
\sqrt{1-\alpha^2/\beta^2-\omega_2^2/\omega_1^2}
\tanh\left(C\xi\right)\right]\right\}.\eea Here we have set the
integration constants $\phi_0$, $\chi_0$ equal to zero. Several
examples, which illustrate the established NR - CSG
correspondence, are considered in the appendix A.

\setcounter{equation}{0}
\section{Finite-Size Effects}

In this section, we will obtain finite-size string solutions,
their images in the (complex) sine-Gordon system, and the leading
corrections to the SS ``$E-\Delta\varphi$''
relation: first for the $R_t\times S^2$ case, then for the SS
string with two angular momenta.

\subsection{Strings on $R_t\times S^2$}

Here we present the solutions of Eq.(\ref{tps2}), when
\bea\nn 0<\frac{\kappa^2}{\omega_1^2}<1,\h
0<\frac{\beta^2\kappa^2}{\alpha^2\omega_1^2}<1\eea
for the two possibilities: $\alpha^2>\beta^2$ and $\alpha^2<\beta^2$.
The first case reduces to the GM string in the limit
$\kappa^2=\omega_1^2$, while the second one corresponds
to the SS solution in $\alpha^2\omega_1^2=\beta^2\kappa^2$ limit.

\subsubsection{The Giant Magnon}

For $\alpha^2>\beta^2$ the solution of Eq.(\ref{tps2}) is given by
\bea\label{FSgms}
&&\cos\theta=\frac{\cos\theta_{min}}{dn\left(C(\xi-\xi_0)|m\right)},\h
\cos\theta_{min}\equiv\sqrt{1-\kappa^2/\omega_1^2},\\ \nn
&&C=\mp\frac{\omega_1\sqrt{1-\beta^2\kappa^2/\alpha^2\omega_1^2}}{\alpha(1-\beta^2/\alpha^2)},
\h
m\equiv\frac{\kappa^2(1-\beta^2/\alpha^2)}{\omega_1^2(1-\beta^2\kappa^2/\alpha^2\omega_1^2)},
\eea where $dn(u|m)$ is one of the elliptic functions and $\xi_0$
is an integration constant. The modulus $m$ is positive.
By using the relation \cite{GR}
\bea\nn dn(u+\mathbf{K}(m)|m)=\frac{\sqrt{1-m}}{dn(u|m)},\eea and
after choosing $C\xi_0=\mathbf{K}$, the above solution can be
rewritten as \bea\label{AFZ}
\cos\theta=\cos\theta_{max}dn\left(C\xi|m\right),\h
\cos\theta_{max}\equiv\sqrt{1-\beta^2\kappa^2/\alpha^2\omega_1^2}
.\eea In this form, (\ref{AFZ}) corresponds to the Arutyunov-Frolov-Zamaklar solution \cite{AFZ06}
(See also \cite{AFGS,KM0803}).

Inserting (\ref{FSgms}) into (\ref{f1s}), one can find \cite{PBMv3}
\bea\nn
f_1=\frac{\beta(\omega_1^2-\kappa^2)}{\omega_1(\alpha^2-\beta^2)}
\left[\xi+\frac{\sqrt{\alpha^2-\beta^2}}{\omega_1^2}
\Pi\left(am(C\xi-\mathbf{K}),-\frac{m}{\kappa^2}\bigg\vert m\right)\right],\eea
where $\Pi$ is the elliptic integral of third
kind. Hence, the string solution is given by \bea\nn
&&W_1=R\sqrt{1-\left(1-\beta^2\kappa^2/\alpha^2\omega_1^2\right)dn^2\left(C\xi|m\right)}\\
\label{fsgms}
&&\times\exp\left\{\frac{i\omega_1}{1-\beta^2/\alpha^2}
\left[(1-\beta^2\kappa^2/\alpha^2\omega_1^2)\tau +
(1-\kappa^2/\omega_1^2) \frac{\beta}{\alpha}\sigma\right]\right.\\
\nn &&+\left.\frac{i\beta(1-\kappa^2/\omega_1^2)}
{\alpha\omega_1\sqrt{1-\beta^2/\alpha^2}}
\Pi\left(am(C\xi-\mathbf{K}),-\frac{m}{\kappa^2}\bigg\vert m\right)\right\},\\
\nn
&&W_2=R\sqrt{1-\beta^2\kappa^2/\alpha^2\omega_1^2}dn\left(C\xi|m\right),\qquad
Z_0=R\exp(i\kappa\tau).\eea

Now, let us see which solution of the SG equation is the image of
(\ref{fsgms}). From Eqs.(\ref{equiv}) and (\ref{equiv1}), we have
\bea\label{sin1}
\sin^2(\phi/2)=\frac{\kappa^2\left(1-\kappa^2/\omega_1^2\right)sn^2\left(C\xi-\mathbf{K}|m\right)}
{M^2\left(1-\beta^2\kappa^2/\alpha^2\omega_1^2\right)dn^2\left(C\xi-\mathbf{K}|m\right)}.\eea
This solution of the CSG system reduces to that of the
SG equation for $M^2=\kappa^2$.
On the other hand, Eq.(\ref{ffi}) with $M^2=\kappa^2$ gives
\bea\nn
\sin^2(\phi/2)=sn^2\left(\mp\frac{\kappa\sqrt{\omega_1^2/\kappa^2-1}}
{\alpha\left(1-\beta^2/\alpha^2\right)}(\xi-\xi_0)\bigg\vert
-\frac{1-\beta^2/\alpha^2}{\omega_1^2/\kappa^2-1}\right).\eea
One can see that the two results match if $M^2=\kappa^2$ and $C\xi_0=\mathbf{K}$
from an identity \cite{GR}
\bea\nn sn\left(u|-m\right)=
\frac{1}{\sqrt{1+m}}\frac{sn\left(u\sqrt{1+m}|m(1+m)^{-1}\right)}
{dn\left(u\sqrt{1+m}|m(1+m)^{-1}\right)}.\eea

In order to find the energy-charge relation for this string
configuration, we need first to compute the conserved quantities.
In accordance with (\ref{cqs}), we have
\bea\nn
\mathcal{E}_s&\equiv&
\frac{2\pi}{\sqrt{\lambda}}E_s=-2\frac{\kappa}{\alpha}
\int_{\theta_{min}}^{\theta_{max}}\frac{d \theta}{\theta'}
=2\frac{\kappa(1-\beta^2/\alpha^2)}
{\omega_1\sqrt{1-\beta^2\kappa^2/\alpha^2\omega_1^2}}\ \mathbf{K}(m),\\\nn
\mathcal{J}&\equiv& \frac{2\pi}{\sqrt{\lambda}}J_1=
2\frac{\alpha\omega_1}{\alpha^2-\beta^2}
\int_{\theta_{min}}^{\theta_{max}}\frac{d\theta}{\theta'}
\left(\sin^2\theta-\frac{\beta^2\kappa^2}{\alpha^2\omega_1^2}\right)
=2\sqrt{1-\beta^2\kappa^2/\alpha^2\omega_1^2} \left[\mathbf{K}(m)
-\mathbf{E}(m)\right],\eea
which leads to
\bea\nn \mathcal{E}_s - \mathcal{J}=
2\sqrt{1-\beta^2\kappa^2/\alpha^2\omega_1}\left[\mathbf{E}(m)-\left(1-\kappa/\omega_1\right)
\frac{1+\beta^2\kappa/\alpha^2\omega_1}
{1-\beta^2\kappa^2/\alpha^2\omega_1^2}\mathbf{K}(m)\right]. \eea
The worldsheet momentum can be expressed as
\bea\nn
p=2\int_{\theta_{min}}^{\theta_{max}}\frac{d
\theta}{\theta'}f'_1=-2\frac{\beta/\alpha}{\sqrt{1-\beta^2\kappa^2/\alpha^2\omega_1^2}}
\left[\frac{\alpha^2}{\beta^2}\Pi\left(1-\frac{\alpha^2}{\beta^2}\bigg\vert m\right)
-\mathbf{K}(m)\right].\eea
In the above expressions, $\mathbf{K}(m)$, $\mathbf{E}(m)$ and
$\Pi(n|m)$ are the complete elliptic integrals.

In terms of new parameters defined by
\bea\nn
\epsilon\equiv 1-m,\h v\equiv -\beta/\alpha,\eea
these expressions can be simplified as follows:
\bea\nn
&&\mathcal{E}_s=2\sqrt{(1-v^2)(1-\epsilon)}\mathbf{K}(1-\epsilon),\h
\mathcal{J}=2\sqrt{\frac{1-v^2}{1-v^2\epsilon}}
\left[\mathbf{K}(1-\epsilon)-\mathbf{E}(1-\epsilon)\right],\\ \nn
&&\mathcal{E}_s-\mathcal{J}=2\sqrt{\frac{1-v^2}{1-v^2\epsilon}}
\left[\mathbf{E}(1-\epsilon)-\left(1-\sqrt{(1-v^2\epsilon)
(1-\epsilon)}\right)\mathbf{K}(1-\epsilon)\right],
\\ \nn &&p=2v\sqrt{\frac{1-v^2\epsilon}{1-v^2}} \left[\frac{1}{v^2}\Pi\left(1-\frac{1}{v^2}|
1-\epsilon\right)-\mathbf{K}(1-\epsilon)\right].\eea We are
interested in the behavior of these quantities in the limit
$\epsilon\to 0$. To establish it, we will use the expansions for
the elliptic functions given in appendix B.

Our approach is as follows. First, we expand $\mathcal{E}_s$,
$\mathcal{J}$ and $p$ about $\epsilon=0$ keeping $v$ independent
of $\epsilon$. Second, we introduce $v(\epsilon)$ according to the
rule \bea\nn v(\epsilon)=v_0(p)+v_1(p)\epsilon +
v_2(p)\epsilon\log(\epsilon)\eea and expand again.
For $p$ to be finite, we find \bea\nn &&v_0(p)=\cos(p/2),\h
v_1(p)=\frac{1}{4}\sin^2(p/2)\cos(p/2)(1-\log(16)),\\ \nn &&
v_2(p)=\frac{1}{4}\sin^2(p/2)\cos(p/2).\eea After that, from the
expansion for  $\mathcal{J}$, we obtain $\epsilon$ as a function
of $\mathcal{J}$ and $p$ \bea\nn \epsilon=16
\exp{\left(-\frac{\mathcal{J}}{\sin(p/2)}-2\right)}.\eea Finally,
using all these in the expansion for $\mathcal{E}_s-\mathcal{J}$,
we derive \bea\nn
\mathcal{E}_s-\mathcal{J}=2\sin(p/2)\left[1-4\sin^2(p/2)
\exp{\left(-\frac{\mathcal{J}}{\sin(p/2)}-2\right)}\right],\eea
which reproduces the leading finite-size effects of the
GM derived in \cite{AFZ06,AFGS,KM0803}.

\subsubsection{The Single Spike}

Now, we are going to consider the second possibility,
namely, $\alpha^2<\beta^2$. This time, the solution of the equation
(\ref{tps2}) can be written as \bea\label{FSsss}
&&\cos\theta=\cos\theta_{max}dn\left(C\xi|m\right),\h
\cos\theta_{max}\equiv\sqrt{1-\kappa^2/\omega_1^2},\\ \nn
&&C=\pm\frac{\alpha\omega_1\sqrt{1-\kappa^2/\omega_1^2}}{\beta^2(1-\alpha^2/\beta^2)},
\h m\equiv\frac{\beta^2/\alpha^2-1}{\omega_1^2/\kappa^2-1}. \eea
Here the new modulus $m$ is positive again.
From (\ref{f1s}), one can find
\bea\nn
f_1=\pm\frac{\beta/\alpha}{\sqrt{1-\kappa^2/\omega_1^2}}
\Pi\left(am(C\xi),\beta^2/\alpha^2-1|m \right) -
\frac{\omega_1}{1-\alpha^2/\beta^2}
\left(\frac{\alpha}{\beta}\sigma+\tau\right).\eea This results in
the following string solution
\bea\nn
&&W_1=R\sqrt{1-\left(1-\kappa^2/\omega_1^2\right)dn^2\left(C\xi|m\right)}
\label{fssss}\\
&&\times\exp\left\{-i\omega_1\frac{\alpha/\beta}{1-\alpha^2/\beta^2}
\left(\sigma+\frac{\alpha}{\beta}\tau\right)\pm
i\frac{\beta/\alpha}{\sqrt{1-\kappa^2/\omega_1^2}}
\Pi\left(am(C\xi),\beta^2/\alpha^2-1|m \right)\right\} \nn,\\ \nn
&&W_2=R\sqrt{1-\kappa^2/\omega_1^2}dn\left(C\xi|m\right),\qquad
Z_0=R\exp(i\kappa\tau).\eea

From Eqs.(\ref{equiv}) and (\ref{equiv1}), the CSG solution corresponding to (\ref{fssss})
\bea
\sin^2(\phi/2)=\frac{\kappa^2}{M^2}sn^2\left(C\xi|m\right)\label{FSssSGs}\eea
becomes that of the SG after fixing $M^2=\kappa^2$.
On the other hand, from (\ref{ffi}) we get
\bea\nn
\sin^2(\phi/2)=sn^2\left(\pm\frac{\kappa\sqrt{\omega_1^2/\kappa^2-1}}
{\alpha(\beta^2/\alpha^2-1)}(\xi-\xi_0)\bigg\vert m\right).\eea
Setting
$\xi_0=0$ and rewriting \bea\nn
\pm\frac{\kappa\sqrt{\omega_1^2/\kappa^2-1}}
{\alpha(\beta^2/\alpha^2-1)}\xi=\pm
\omega_1\frac{\alpha\sqrt{1-\kappa^2/\omega_1^2}}{\beta(1-\alpha^2/\beta^2)}
\left(\frac{\alpha}{\beta}\sigma+\tau\right)=C\xi,\eea we find
agreement with (\ref{FSssSGs}) if $M^2=\kappa^2$.

Next, let us compute the conserved quantities for the present
string solution. By using (\ref{cqs}), we receive
\bea\nn
\mathcal{E}_s&=&2\frac{\kappa}{\alpha}
\int_{\theta_{min}}^{\theta_{max}}\frac{d \theta}{\theta'}
=2\frac{\kappa(\beta^2/\alpha^2-1)}
{\omega_1\sqrt{1-\kappa^2/\omega_1^2}}\mathbf{K}(m),\nn\\
\mathcal{J}&=&
\frac{2}{\alpha} \int_{\theta_{min}}^{\theta_{max}}\frac{d
\theta}{\theta'}\sin^2\theta\left(\beta f'_1+\omega_1\right)
=2\sqrt{1-\kappa^2/\omega_1^2} \left[\mathbf{E}(m)
-\frac{1-\beta^2\kappa^2/\alpha^2\omega_1^2}{1-\kappa^2/\omega_1^2}\mathbf{K}
(m)\right].\nn
\eea
In addition, we compute $\Delta\varphi_1$
\bea\nn
\Delta\varphi\equiv\Delta\varphi_1=2\int_{\theta_{min}}^{\theta_{max}}\frac{d
\theta}{\theta'}f'_1=-2\frac{\beta/\alpha}{\sqrt{1-\kappa^2/\omega_1^2}}
\left[\Pi\left(1-\frac{\beta^2}{\alpha^2}|
m\right)-\mathbf{K}(m)\right].\eea
Defining parameters
\bea\nn
\epsilon\equiv 1-m,\h
v\equiv \beta/\alpha,\eea we can rewrite these as
\bea\nn
&&\mathcal{E}_s=2\sqrt{(v^2-1)(1-\epsilon)}\mathbf{K}(1-\epsilon),\h
\mathcal{J}=2\sqrt{\frac{v^2-1}{v^2-\epsilon}}
\left[\mathbf{E}(1-\epsilon)-\epsilon\mathbf{K}(1-\epsilon)\right],\\
\nn &&\Delta\varphi=-2v\sqrt{\frac{v^2-\epsilon}{v^2-1}}
\left[\Pi\left(1-v^2|
1-\epsilon\right)-\mathbf{K}(1-\epsilon)\right]\\ \nn
&&\mathcal{E}_s-\Delta\varphi=2v\sqrt{\frac{v^2-\epsilon}{v^2-1}}
\left[\Pi\left(1-v^2|
1-\epsilon\right)-\left(1-\frac{(v^2-1)\sqrt{1-\epsilon}}{v\sqrt{v^2-\epsilon}}\right)
\mathbf{K}(1-\epsilon)\right].\eea

We proceed as in the GM case to find the $\epsilon$ expansion of these quantities.
The difference is that now we impose $\mathcal{J}$ to remain finite, which gives \bea\nn
\mathcal{J}=2\sqrt{1-\frac{1}{v_0^2}},\h
v_1=\frac{(v_0^2-1)\left[v_0^2(1+\log(16))-2\right]}{4v_0},\h
v_2=-\frac{v_0(v_0^2-1)}{4}.\eea From the expansion for
$\Delta\varphi$, we obtain $\epsilon$ as a function of
$\Delta\varphi$ and $\mathcal{J}$
\bea\nn \epsilon=16
\exp\left(-\frac{\sqrt{4-\mathcal{J}^2}}{\mathcal{J}}\left[\Delta\varphi+
\arcsin\left(\frac{\mathcal{J}}{2}\sqrt{4-\mathcal{J}^2}\right)\right]
\right).\eea Using
these results in the expansion for $\mathcal{E}_s-\Delta\varphi$,
one can see that the divergent terms cancel each other for
$\mathcal{J}^2<2$ and the finite result is
\bea\nn E_s-\frac{\sqrt{\lambda}}{2\pi}\Delta\varphi &=& \frac{\sqrt{\lambda}}{\pi}
\left[\frac{1}{2}\arcsin\left(\frac{\mathcal{J}}{2}\sqrt{4-\mathcal{J}^2}\right)
+\frac{\mathcal{J}^3}{16\sqrt{4-\mathcal{J}^2}}
\ \epsilon\right]\\
\label{ssS2c} &=&\frac{\sqrt{\lambda}}{\pi}
\left[\frac{p}{2}+4\sin^2\frac{p}{2}\tan\frac{p}{2}
\exp\left(-\frac{\Delta\varphi+ p}
{\tan\frac{p}{2}}\right)\right],\eea where we used the
identification \cite{IK07,IKSV07} \bea\nn
\arcsin\left(\mathcal{J}/2\right)=\frac{p}{2}=\bar{\theta}
=\pi/2-\arcsin\frac{\kappa}{\omega_1}.\eea
This is the
leading finite-size correction to the SS ``$E-\Delta\varphi$''
relation. Let us also note that to the leading order, the length
$L$ of this SS string can be computed to be \bea\nn
L=\frac{\alpha}{\kappa}\left(\Delta\varphi+ p\right).\eea

\subsection{Strings on $R_t\times S^3$}

In the case $C_2=0$, Eq.(\ref{tsol}) can be written as
\bea\label{tS3eq}
(\cos\theta)'=\mp\frac{\alpha\sqrt{\omega_1^2-\omega_2^2}}{\alpha^2-\beta^2}
\sqrt{(z_+^2-\cos^2\theta)(\cos^2\theta-z_-^2)},\eea where \bea\nn
&&z^2_\pm=\frac{1}{2(1-\frac{\omega_2^2}{\omega_1^2})}
\left\{y_1+y_2-\frac{\omega_2^2}{\omega_1^2}
\pm\sqrt{(y_1-y_2)^2-\left[2\left(y_1+y_2-2y_1
y_2\right)-\frac{\omega_2^2}{\omega_1^2}\right]
\frac{\omega_2^2}{\omega_1^2}}\right\}, \\ \nn
&&y_1=1-\kappa^2/\omega_1^2,\h
y_2=1-\beta^2\kappa^2/\alpha^2\omega_1^2 .\eea The solution of
(\ref{tS3eq}) can be obtained as \bea\label{tS3sol} \cos\theta=z_+
dn\left(C\xi|m\right),\h
C=\mp\frac{\alpha\sqrt{\omega_1^2-\omega_2^2}}{\alpha^2-\beta^2}
z_+,\h m\equiv 1-z^2_-/z^2_+ .\eea The solutions of Eqs.(\ref{f1s}) and (\ref{f2s})
now read \bea\nn
&&f_1=\frac{2\beta/\alpha}{z_+\sqrt{1-\omega_2^2/\omega_1^2}}
\left[F\left(am(C\xi)|m\right) -
\frac{\kappa^2/\omega_1^2}{1-z^2_+}\Pi\left(am(C\xi),-\frac{z^2_+
-z^2_-}{1-z^2_+}|m\right)\right],\\ \nn
&&f_2=\frac{2\beta\omega_2/\alpha\omega_1}{z_+\sqrt{1-\omega_2^2/\omega_1^2}}
F\left(am(C\xi)|m\right).\eea Therefore, the full string solution
is given by \bea\nn &&Z_0=R\exp(i\kappa\tau),\\
\nn &&W_1=R\sqrt{1-z^2_+
dn^2\left(C\xi|m\right)}\exp\left\{i\omega_1\tau +
\frac{2i\beta/\alpha}{z_+\sqrt{1-\omega_2^2/\omega_1^2}}\right.\\
\nn &&\times\left.\left[F\left(am(C\xi)|m\right) -
\frac{\kappa^2/\omega_1^2}{1-z^2_+}\Pi\left(am(C\xi),-\frac{z^2_+
-z^2_-}{1-z^2_+}|m\right)\right]\right\} ,\\
\label{fssS3} &&W_2=Rz_+
dn\left(C\xi|m\right)\exp\left\{i\omega_2\tau
+\frac{2i\beta\omega_2/\alpha\omega_1}
{z_+\sqrt{1-\omega_2^2/\omega_1^2}}
F\left(am(C\xi)|m\right)\right\} .\eea We note that (\ref{fssS3})
contains both cases: $\alpha^2>\beta^2$ for the GM and
$\alpha^2<\beta^2$ for the SS.

To find the CSG solution related to (\ref{fssS3}), we insert
(\ref{tS3sol}) into (\ref{equiv}) and (\ref{equiv1}) to get
\bea\label{fisS3}
\sin^2(\phi/2)=\frac{\omega_1^2/M^2}{\beta^2/\alpha^2-1}
\left[\left(1-\kappa^2/\omega_1^2\right) -
\left(1-\omega_2^2/\omega_1^2\right)\left(z^2_+ cn^2(C\xi|m) +
z^2_- sn^2(C\xi|m)\right)\right].\eea After that, we use
(\ref{fisS3}) in (\ref{kfi}) and integrate. The result is
\bea\label{chiS3} \chi=\frac{A}{\beta}(\beta\sigma+\alpha\tau) -
C_\chi(\alpha\sigma+\beta\tau) + \frac{C_\chi}{C
D}\Pi\left(am(C\xi),n|m\right),\eea where $A/\beta$ and $C_\chi$
are given in (\ref{equiv2}), $C_2=0$, and
\bea\nn D=\frac{\omega_1^2/M^2}{\beta^2/\alpha^2-1}
\left[\left(1-\kappa^2/\omega_1^2\right) -
\left(1-\omega_2^2/\omega_1^2\right)z^2_+\right],\h
n=\frac{\left(1-\omega_2^2/\omega_1^2\right)
(z^2_+-z^2_-)}{\left(1-\kappa^2/\omega_1^2\right) -
\left(1-\omega_2^2/\omega_1^2\right)z^2_+}.\eea Hence for the
present case, the CSG field $\psi=\sin(\phi/2)\exp(i\chi/2)$ is
defined by (\ref{fisS3}) and (\ref{chiS3}).

In a recent paper \cite{HS08}, the finite-size effects for
dyonic GM have been considered and the leading order correction to
the $\mathcal{E}_s-\mathcal{J}_1$ relation has been found to be
\bea\nn
\mathcal{E}_s-\mathcal{J}_1&=&\sqrt{\mathcal{J}^2_2 +
4\sin^2(p_1/2)}\\ \nn &-&
8\frac{\sin^3(p_1/2)}{\cosh(\tilde{\theta}/2)}
\exp\left[-\frac{2\sin^2(p_1/2)\cosh^2(\tilde{\theta}/2)}
{\sin^2(p_1/2) +
\sinh^2(\tilde{\theta}/2)}\left(\frac{\mathcal{J}_1/2}
{\sin(p_1/2)\cosh(\tilde{\theta}/2)}+1\right)\right],\eea
where
\bea\nn \cosh(\tilde{\theta}/2)=\frac{\sqrt{(\mathcal{J}_2/2)^2 +
\sin^2(p_1/2)}}{\sin(p_1/2)}.\eea

Our concern here is the SS for the case
$\alpha^2<\beta^2$  with two angular
momenta $\mathcal{J}_1$ and $\mathcal{J}_2$. The computation of
the conserved quantities (\ref{cqs}) and $\Delta\varphi_1$ now
gives
\bea\nn &&\mathcal{E}_s
=\frac{2\kappa(\beta^2/\alpha^2-1)}
{\omega_1\sqrt{1-\omega_2^2/\omega_1^2}z_+}\mathbf{K}
\left(1-z^2_-/z^2_+\right), \\ \nn &&\mathcal{J}_1= \frac{2
z_+}{\sqrt{1-\omega_2^2/\omega_1^2}} \left[\mathbf{E}
\left(1-z^2_-/z^2_+\right)
-\frac{1-\beta^2\kappa^2/\alpha^2\omega_1^2}{z^2_+}\mathbf{K}
\left(1-z^2_-/z^2_+\right)\right], \\ \nn &&\mathcal{J}_2=
-\frac{2 z_+ \omega_2/\omega_1
}{\sqrt{1-\omega_2^2/\omega_1^2}}\mathbf{E}
\left(1-z^2_-/z^2_+\right), \\ \nn
&&\Delta\varphi=
-\frac{2\beta/\alpha}{\sqrt{1-\omega_2^2/\omega_1^2}z_+}
\left[\frac{\kappa^2/\omega_1^2}{1-z^2_+}\Pi\left(-\frac{z^2_+ -
z^2_-}{1-z^2_+}|1-z^2_-/z^2_+\right) -\mathbf{K}
\left(1-z^2_-/z^2_+\right)\right].\eea Our next step is to
introduce the new parameters \bea\nn \epsilon\equiv z^2_-/z^2_+,\h
v\equiv\beta/\alpha,\h u\equiv\omega^2_2/\omega^2_1 ,\eea and to rewrite
$\mathcal{E}_s$, $\mathcal{J}_1$, $\mathcal{J}_2$, $\Delta\varphi$
in the form \bea\nn &&\mathcal{E}_s=2 K_e
\mathbf{K}\left(1-\epsilon\right), \\ \label{ncoeff}
&&\mathcal{J}_1=2 K_{11}\left[\mathbf{E}\left(1-\epsilon\right)
-K_{12}\mathbf{K}\left(1-\epsilon\right)\right], \\ \nn
&&\mathcal{J}_2=2 K_2 \mathbf{E}\left(1-\epsilon\right), \\ \nn
&&\Delta\varphi=2 K_{\varphi 1} \left[K_{\varphi
2}\Pi\left(K_{\varphi 3}|1-\epsilon\right) -\mathbf{K}
\left(1-\epsilon\right)\right].\eea
The explicit $\epsilon$-expansions of the coefficients $K_e,\ldots,K_{\varphi 3}$
are given as functions of $u$ and $v$ in
appendix B.

We also need to consider the $\epsilon$-expansion for $u$ and $v$ as follows:
\bea\nn v(\epsilon)=v_0+v_1\epsilon +
v_2\epsilon\log(\epsilon),\h u(\epsilon)=u_0+u_1\epsilon +
u_2\epsilon\log(\epsilon).\eea
The coefficients can be determined by the condition that
$\mathcal{J}_1$ and $\mathcal{J}_2$ should be finite,
\bea\label{j1j2}
&&v_0=\frac{2\mathcal{J}_1}{\sqrt{\left(\mathcal{J}_1^2-\mathcal{J}_2^2\right)
\left[4-\left(\mathcal{J}_1^2-\mathcal{J}_2^2\right)\right]}},\h
u_0=\frac{\mathcal{J}_2^2}{\mathcal{J}_1^2},
\\ \nn
&&v_1=\frac{(1-u_0)v_0^2-1}{4(u_0-1)(v_0^2-1)v_0}
\left\{(u_0-1)v_0^4(1+\log(16))-2\right.\\ \label{uvs} &&+ \left.
v_0^2\left[3+\log(16)+u_0(\log(4096)-5)\right]\right\},
\\ \nn
&&v_2=-\frac{v_0\left[1-(1-u_0)v_0^2\right]\left[1+3u_0-(1-u_0)v_0^2\right]}
{4(1-u_0)(v_0^2-1)},\\ \nn
&&u_1=\frac{u_0\left[1-(1-u_0)v_0^2\right]\log(16)}{v_0^2-1},\h
u_2=-\frac{u_0\left[1-(1-u_0)v_0^2\right]}{v_0^2-1}. \eea
The parameter $\epsilon$ can be obtained from $\Delta\varphi$ as follows:
\bea\label{es} \epsilon=16
\exp\left(-\frac{\sqrt{(1-u_0)v_0^2-1}}{v_0^2-1}\left[\Delta\varphi +
\arcsin\left(\frac{2\sqrt{(1-u_0)v_0^2-1}}
{(1-u_0)v_0^2}\right)\right]\right).\eea
From Eqs.(\ref{j1j2}), (\ref{uvs}) and (\ref{es}), $\mathcal{E}_s-\Delta\varphi$ can be
derived as
\bea\label{DiffJ12}\mathcal{E}_s-\Delta\varphi&=&
\arcsin N(\mathcal{J}_1,\mathcal{J}_2)
+ 2\left(\mathcal{J}_1^2-\mathcal{J}_2^2\right)
\sqrt{\frac{4}{\left[4-\left(\mathcal{J}_1^2-\mathcal{J}_2^2\right)\right]}-1}\\
&\times&\exp\left[-\frac{2\left(\mathcal{J}_1^2-\mathcal{J}_2^2\right)
N(\mathcal{J}_1,\mathcal{J}_2)}
{\left(\mathcal{J}_1^2-\mathcal{J}_2^2\right)^2 +
4\mathcal{J}_2^2}\left[\Delta\varphi
+\arcsin N(\mathcal{J}_1,\mathcal{J}_2)\right]\right],\\
N(\mathcal{J}_1,\mathcal{J}_2)&\equiv&
\frac{1}{2}\left[4-\left(\mathcal{J}_1^2-\mathcal{J}_2^2\right)\right]
\sqrt{\frac{4}{\left[4-\left(\mathcal{J}_1^2-\mathcal{J}_2^2\right)\right]}-1}.\eea
Here $\mathcal{J}_1^2-\mathcal{J}_2^2<2$ is assumed.
Finally, by using the SS relation between the angular momenta
\bea\nn
\mathcal{J}_1=\sqrt{\mathcal{J}_2^2+4\sin^2(p/2)},\eea
we obtain ($-\pi/2\le p\le\pi/2$)
\bea\label{ssS3c}
E_s-\frac{\sqrt{\lambda}}{2\pi}\Delta\varphi=
\frac{\sqrt{\lambda}}{\pi}
\left[\frac{p}{2}+4\sin^2\frac{p}{2}\tan\frac{p}{2}
\exp\left(-\frac{\tan\frac{p}{2}(\Delta\varphi+ p)}
{\tan^2\frac{p}{2} + \mathcal{J}_2^2 \csc^2p}\right)\right].\eea
This is our final result for the leading
finite-size correction to the ``$E-\Delta\varphi$'' relation for
the SS string with two angular momenta. It is obvious that for
$\mathcal{J}_2=0$ (\ref{ssS3c}) reduces to (\ref{ssS2c}) as it
should be.

\section{Concluding Remarks}

In this paper, by using the reduction of the string dynamics on
$R_t\times S^3$ to the NR integrable system, we gave an
explicit mapping connecting the parameters of {\it all} string
solutions described by this dynamical system and the parameters in
the corresponding solutions of the complex sine-Gordon integrable
model. In the framework of this NR approach, we found finite-size
string solutions, their images in the (complex) sine-Gordon
system, and the leading finite-size corrections to the
single spike ``$E-\Delta\varphi$'' relation: both for strings on
$R_t\times S^2$ and $R_t\times S^3$ backgrounds.
It is an important open question to compare our results on the finite-size
effects of the single spikes with the L\"uscher corrections obtained
from the exact $S$-matrices.
The classical scattering amplitude computed in \cite{IKSV07} turns out
to be not sufficient since its complete pole structure is not clear.
We hope that our results in the classical limit provide a clue to figure out
the exact quantum $S$-matrix for the single spikes.

The GM and SS energy-charge relations for strings on $R_t\times
S^5$ are already known for the infinite case \cite{KRT06,DimRas}.
This opens a possibility to find the finite-size effects for
such generalized string configurations. We are convinced that the
NR approach will be effective in this case too.
An evident direction of further development is to consider string
configurations in the $AdS$ part of the full $AdS_5\times S^5$
background. It is known \cite{ART} that the corresponding
integrable system will be again of NR-type, but with indefinite
signature.
Another interesting case is strings with nonzero spins on both
$AdS_5$ and $S^5$ part of the target space. In this case, we will
have two NR-type systems. While the equations of motion for them
will decouple, the variables of the two NR systems will be mixed
in the constraints \cite{ART}. Thus, a new kind of problem will
appear. Nevertheless, there may exist string configurations for
which this problem is solvable as found in \cite{DimRas} for
example.

\section*{Acknowledgements}
We thank R. C. Rashkov for comments on the draft. This work was
supported in part by KRF-2007-313-C00150 (CA), by NSFB VU-F-201/06
(PB), and by the Brain Pool program from the Korean Federation of
Science and Technology (2007-1822-1-1).

\section*{Appendices}
\def\theequation{A.\arabic{equation}}
\setcounter{equation}{0}
\begin{appendix}

\section{Relation between the NR system and CSG}

\subsection{Explicit Relations between the Parameters}

In the general case, the relation between the parameters in the
solutions of the NR and CSG integrable systems is given by \bea\nn
K^2=R^2M^2,\h C_\phi= \frac{2}{\alpha^2-\beta^2}\left\{3M^2-
2\left[\kappa^2
+\frac{(\kappa^2-\omega_1^2)-\omega_2^2}{1-\beta^2/\alpha^2}\right]\right\},\eea
\bea\nn &&\frac{1}{4}M^4(\alpha^2-\beta^2)\frac{A^2}{\beta^2}=
M^4\left(M^2-\kappa^2+\frac{\omega_2^2}{1-\beta^2/\alpha^2}\right)
\\ \nn &&-\left(\frac{\kappa^2-\omega_1^2}{1-\beta^2/\alpha^2}\right)
\left\{M^4 +
\left[M^2-\left(\frac{\kappa^2-\omega_1^2}{1-\beta^2/\alpha^2}\right)\right]
\left(\frac{\kappa^2-\omega_1^2}{1-\beta^2/\alpha^2}\right)\right.\\
\label{equiv2}
&&-\left.\left[2M^2-\left(\frac{\kappa^2-\omega_1^2}{1-\beta^2/\alpha^2}\right)\right]
\left(\kappa^2-\frac{\omega_2^2}{1-\beta^2/\alpha^2}\right)\right\}
\\ \nn &&-\frac{(\omega_1^2-\omega_2^2)}{\omega_1^2\left(1-\beta^2/\alpha^2\right)^3}
\biggl\{\left[M^2\left(1-\beta^2/\alpha^2\right) - \kappa^2\right]
(\omega_1^2-\omega_2^2)\check{C}_2^2\\ \nn
&&-\left[M^2\left(1-\beta^2/\alpha^2\right) -
(\kappa^2-\omega_1^2)\right]
\left[2\frac{\beta}{\alpha}\omega_2\kappa^2\check{C}_2 +
\left(\kappa^2-\omega_1^2\right)
\left(\frac{\beta^2}{\alpha^2}\kappa^2-\omega_1^2\right)\right]\biggr\},\eea
\bea \nn &&\frac{1}{4}M^4(\alpha^2-\beta^2)C_\chi^2=
-\left(\frac{\kappa^2-\omega_1^2}{1-\beta^2/\alpha^2}\right)^2
\left[\kappa^2
-\frac{(\kappa^2-\omega_1^2)+\omega_2^2}{1-\beta^2/\alpha^2}\right]
\\ \nn &&+\frac{(\omega_1^2-\omega_2^2)}{\omega_1^2\left(1-\beta^2/\alpha^2\right)^3}
\left\{\kappa^2 (\omega_1^2-\omega_2^2)\check{C}_2^2-(\kappa^2-\omega_1^2)
\left[2\frac{\beta}{\alpha}\omega_2\kappa^2\check{C}_2 +
\left(\kappa^2-\omega_1^2\right)
\left(\frac{\beta^2}{\alpha^2}\kappa^2-\omega_1^2\right)\right]\right\}
,\eea where $\check{C}_2=C_2/\alpha$. Thus, we have expressed the CSG
parameters $C_\phi$, $A$ and $C_\chi$ through the NR parameters
$\alpha$, $\beta$, $\kappa$, $\omega_1$, $\omega_2$, $C_2$. The
mass parameter $M$ remains free.

Let us consider several
examples, which illustrate the established NR - CSG
correspondence. We are interested in the GM and SS
configurations on $R_t\times S^2$ and $R_t\times S^3$.
From the NR-system viewpoint, we have to set $C_2=0$ in
(\ref{tsol}), (\ref{f1s}) and (\ref{f2s}) for the GM and SS string
solutions. This condition is to require one of the
turning points, where $\theta'=0$, to lay on the equator of the
sphere, i.e. $\theta=\pi/2$ \cite{KRT06}.

\subsection{On $R_t\times S^2$}
We begin with the $R_t\times S^2$ case, when $C_2=\omega_2=0$ and
$\theta'$ in (\ref{tsol}) takes the form \bea\label{tps2}
\theta'=\frac{\pm\alpha\omega_1}{(\alpha^2-\beta^2)\sin\theta}
\sqrt{\left(\frac{\beta^2\kappa^2}{\alpha^2\omega_1^2}-\sin^2\theta\right)
\left(\sin^2\theta-\frac{\kappa^2}{\omega_1^2}\right)}.\eea

\subsubsection{The Giant Magnon}

The GM solution corresponds to $\kappa^2=\omega_1^2$ with $\alpha^2>\beta^2$, which is
given by
\bea\nn \cos\theta=
\frac{\sqrt{1-\beta^2/\alpha^2}}{\cosh\left(\omega_1\frac{\sigma+\tau\beta/\alpha}
{\sqrt{1-\beta^2/\alpha^2}}\right)}.
\label{gmsol}\eea
From \ref{f1s}), one finds $f_2=0$ and
\bea\nn
f_1=\arctan\left[\frac{\alpha}{\beta}\sqrt{1-\beta^2/\alpha^2}
\tanh\left(\omega_1\frac{\sigma+\tau\beta/\alpha}
{\sqrt{1-\beta^2/\alpha^2}}\right)\right].\eea For
$R^2=M^2=\omega_1^2=1$, $\beta/\alpha=-\sin\theta_0$, this string
solution coincides with the Hofman-Maldacena solution \cite{HM06},
and is equivalent to the solution in \cite{CDO06} for $R_t\times
S^2$ after the identification $W_1=Z_1\exp(i\pi/2)$, $W_2=Z_2$.
Now, the parameters in (\ref{GMp}) take the values \bea\nn
&&K^2=R^2\omega_1^2=1,\h M^2=\omega_1^2=1,\\ \nn
&&C_\phi=\frac{2\omega_1^2}{\alpha^2-\beta^2}
=\frac{2}{\alpha^2\cos^2\theta_0},\h A= C_{\chi}=0,\eea
and from Eqs.(\ref{ffi}),(\ref{equiv}) and (\ref{equiv1})) the corresponding
SG solution becomes
\bea\nn
\sin(\phi/2)=\frac{1}{\cosh\left(\frac{\sigma-\tau\sin\theta_0}
{\cos\theta_0}-\eta_0\right)}.\eea
This can be also obtained from
(\ref{S3SG}) by setting $\omega_2=0$.

However, for $M^2>\omega_1^2=\kappa^2$, we have \bea\nn
A=2\beta\sqrt{\frac{M^2-\omega_1^2}{\alpha^2-\beta^2}}\ne 0.\eea
This case is related to the CSG system instead of the SG one. It is interesting
to find the CSG solution associated with it. Using (\ref{ffi})
again, we find \bea\nn
\sin(\phi/2)=\frac{\omega_1}{M\cosh\left(\omega_1\frac{\sigma+\tau\beta/\alpha}
{\sqrt{1-\beta^2/\alpha^2}}-\eta_0\right)},\h \chi=2
\sqrt{\frac{M^2-\omega_1^2}{1-\beta^2/\alpha^2}}
\left(\frac{\beta}{\alpha}\sigma+\tau\right).\eea

\subsubsection{The Single Spike}

The SS solution corresponds to $\beta^2\kappa^2=\alpha^2\omega_1^2$.
In this case, the expressions
for $\theta$ and $f_1$ are
\bea\nn \cos\theta=
\frac{\sqrt{1-\alpha^2/\beta^2}}{\cosh\left(C\xi\right)},\quad
f_1=-\omega_1(\sigma\alpha/\beta+\tau)+\arctan\left[\frac{\beta}{\alpha}\sqrt{1-\alpha^2/\beta^2}
\tanh\left(C\xi\right)\right],\eea
and the corresponding string solution is
\bea\nn
W_1&=&R\sqrt{1-\frac{1-\alpha^2/\beta^2}{\cosh^2\left(C\xi\right)}}
\exp{\left\{-i\omega_1\sigma\alpha/\beta +
i\arctan\left[\frac{\beta}{\alpha}\sqrt{1-\alpha^2/\beta^2}
\tanh\left(C\xi\right)\right]\right\}},\\
\nn W_2&=&
\frac{R\sqrt{1-\alpha^2/\beta^2}}{\cosh\left(C\xi\right)},\qquad
Z_0=R\exp\left(i\frac{\alpha}{\beta}\omega_1\tau\right),\eea
where we used a short notation
\bea\nn
C\xi\equiv\omega_1\frac{\alpha}{\beta}\frac{\sigma\alpha/\beta+\tau}
{\sqrt{1-\alpha^2/\beta^2}}.
\eea

The ``dual'' SG solution can be obtained from (\ref{phis}) by
setting $\omega_2=0$. If we choose $R=1$,
$\alpha/\beta=\sin\theta_1$, $\omega_1=-\cot\theta_1$, $\beta=1$,
the SS solution on $R_t\times S^2$ in \cite{IK07} is reproduced.

\subsection{On $R_t\times S^3$}

\subsubsection{The Giant Magnon}

Let us continue with the $R_t\times S^3$ case, when $C_2=0$,
$\omega_2\ne 0$. First, we would like to establish the
correspondence between the dyonic GM string
solution \cite{KRT06}
$(\kappa^2=\omega_1^2)$ to those found in \cite{CDO06} \bea\nn
&&Z_1=\frac{1}{\sqrt{1+k^2}} \left\{\tanh\left[\cos\alpha^D
\left(\sigma\sqrt{1+k^2\cos^2\alpha^D}
-k\tau\cos\alpha^D\right)\right]-ik\right\}\exp(i\tau),\\
\nn &&Z_2=\frac{1}{\sqrt{1+k^2}}\frac{\exp\left[i\sin\alpha^D
\left(\tau\sqrt{1+k^2\cos^2\alpha^D}-k\sigma\cos\alpha^D\right)\right]}
{\cosh\left[\cos\alpha^D \left(\sigma\sqrt{1+k^2\cos^2\alpha^D}
-k\tau\cos\alpha^D\right)\right]},\eea
where the
parameter $k$ is related to the soliton rapidity $\hat{\theta}$
through the equality \bea\nn
k=\frac{\sinh\hat{\theta}}{\cos\alpha^D},\eea and $\alpha^D$ determines the $U(1)$ charge
carried by the CSG soliton \cite{CDO06}.

The solutions of Eqs.(\ref{tsol}), (\ref{f1s}) and
(\ref{f2s}) are given by \bea\nn
&&\cos\theta=\frac{\cos\theta_0}{\cosh\left(C\xi\right)},\h
f_1=\arctan\left[\cot\theta_0\tanh(C\xi)\right],
\h f_2=\frac{\beta\omega_2}{\alpha^2-\beta^2}\xi,
\\ \nn && \sin^2\theta_0\equiv\frac{\beta^2\omega_1^2}{\alpha^2(\omega_1^2-\omega_2^2)},\h
C\equiv\frac{\alpha\sqrt{\omega_1^2-\omega_2^2}}
{\alpha^2-\beta^2}\cos\theta_0.\eea
Then, the comparison shows that the two solutions are equivalent
if \bea\nn
&&Z_1\exp(i\pi/2)=W_1=R\sin\theta\exp\left[i\left(\omega_1\tau+f_1\right)\right],
\\ \nn
&&Z_2=W_2=R\cos\theta\exp\left[i\left(\omega_2\tau+f_2\right)\right],
\\ \nn &&R=\kappa=\omega_1=1,\h
\alpha=\cos\alpha^D\sqrt{1+k^2\cos^2\alpha^D},\\ \nn
&&\beta=-k\cos^2\alpha^D,\h\omega_2=
\frac{\sin\alpha^D}{\sqrt{1+k^2\cos^2\alpha^D}}.\eea As a
consequence, the CSG parameters in (\ref{GMp}) reduce to \bea\nn
C_\phi= \frac{2}{\cos^2\alpha^D}\left(1+
2\sin^2\alpha^D\right)],\h A=k\sin(2\alpha^D), \h
C_{\chi}=0,\h K^2=1.\eea

\subsubsection{The Single Spike}

Now, let us turn to the SS solutions on $R_t\times S^3$ as
described by the NR integrable system \cite{BobR07}. By using the
SS-condition $\beta^2\kappa^2=\alpha^2\omega_1^2$ in (\ref{tsol})
one derives \bea\nn
\theta'=\frac{\alpha\sqrt{\omega_1^2-\omega_2^2}}{\alpha^2-\beta^2}
\frac{\cos\theta}{\sin\theta}
\sqrt{\sin^2\theta-\frac{\alpha^2\omega_1^2}{\beta^2(\omega_1^2-\omega_2^2)}},\eea
whose solution is given by \bea\nn \cos\theta =
\frac{\sqrt{\left(1-\alpha^2/\beta^2\right)\omega_1^2-
\omega_2^2}}
{\sqrt{\omega_1^2-\omega_2^2}\cosh(C\xi)},\quad C\xi\equiv
\sqrt{\omega_1^2-\frac{\omega_2^2}{1-\alpha^2/\beta^2}}
\frac{\alpha(\sigma\alpha/\beta+\tau)}
{\sqrt{\beta^2-\alpha^2}}.\eea
By using
(\ref{f1s}), (\ref{f2s}), one finds the following expressions for the string embedding
coordinates $\varphi_j=\omega_j\tau+f_j$
\bea\nn \varphi_1&=&-\omega_1\sigma\alpha/\beta +
\arctan\left\{\frac{\beta}{\alpha\omega_1}\sqrt{\left(1-\frac{\alpha^2}{\beta^2}\right)
\left(\omega_1^2-\frac{\omega_2^2}{1-\alpha^2/\beta^2}\right)}
\tanh(C\xi)\right\}\\
\nn \varphi_2&=&-\omega_2\frac{\alpha(\sigma
+\tau\alpha/\beta)}{\beta(1-\alpha^2/\beta^2)}.\eea Comparing the
above results with the SS string solution given in (4.1) - (4.7)
of \cite{IKSV07}, we see that the two solutions coincide for
\bea\label{chp} R=1,\h
\sin\theta_1=-\frac{1}{\sqrt{\omega_1^2-\omega_2^2}}, \h
\sin\gamma_1=\frac{\omega_2}{\omega_1},\h
\omega_1=-\frac{\beta}{\alpha}.\eea
From (\ref{SSp}), the CSG parameters are
\bea\nn &&C_\phi=
\frac{2}{\beta^2\left(1-\sin^2\theta_1\cos^2\gamma_1\right)}
\left[4-3M^2 +
\frac{2\cos^4\gamma_1}{\sin^2\gamma_1\left(1-\sin^2\theta_1\cos^2\gamma_1\right)}\right],
\h K^2=M^2,\\ \nn
&&A=\frac{M^2-1}{M^2\sqrt{1-\sin^2\theta_1\cos^2\gamma_1}}
\sqrt{\frac{\cos^4\gamma_1}{\sin^2\gamma_1\left(1-\sin^2\theta_1\cos^2\gamma_1\right)}
-M^2},\\ \nn &&C_\chi=
-\frac{2\sin\gamma_1}{M^2\beta\left(1-\sin^2\theta_1\cos^2\gamma_1\right)}.\eea
Comparing (\ref{chp}) with (\ref{asszo}), one sees that the
solution found in \cite{IKSV07} corresponds actually to $M^2=1$
which leads to $A_{SS}=0$. Hence, the ``dual'' CSG solution is of
the type (\ref{CSGsol}).

\section{$\epsilon$-Expansions}

We use the following expansions for the elliptic functions \bea\nn
&&\mathbf{K}(1-\epsilon)\propto
-\frac{1}{2}\log\epsilon\left(1+O(\epsilon)\right)+\log(4)\left(1+O(\epsilon)\right),
\\ \nn
&&\mathbf{E}(1-\epsilon)\propto 1-\epsilon\left(\frac{1}{4}-\log(2)\right)
\left(1+O(\epsilon)\right)-\frac{\epsilon}{4}\log\epsilon\left(1+O(\epsilon)\right)
\\ \nn &&\Pi(n|1-\epsilon)\propto
\frac{\log\epsilon}{2(n-1)}\left(1+O(\epsilon)\right) +
\frac{\sqrt{n}\log\left(\frac{1+\sqrt{n}}{1-\sqrt{n}}\right)-\log(16)}{2(n-1)}
\left(1+O(\epsilon)\right).\eea

The expansions for the coefficients in (\ref{ncoeff}) are given by
\bea\nn &&K_e \propto
\frac{v^2-1}{\sqrt{v^2(1-u)-1}}-\frac{v^2(1-u)^2-1}{2
\sqrt{v^2(1-u)-1}(1-u)}\ \epsilon, \\ \nn &&K_{11}\propto
\sqrt{\frac{v^2(1-u)-1}{v^2(1-u)^2}} -
\frac{\sqrt{v^2(1-u)-1}\left(1+v^2(2u-1)\right)}{2v^3(v^2-1)(1-u)^2}\ \epsilon,
\\ \nn &&K_{12}\propto
\left(1-\frac{v^2u}{v^2-1}\right)\epsilon,
\\ \nn &&K_{2}\propto
-\sqrt{\frac{\left(v^2(1-u)-1\right)u}{v^2(1-u)^2}}
+\frac{\sqrt{\frac{\left(v^2(1-u)-1\right)u}{v^2(1-u)^2}}\left(1+v^2(2u-1)\right)}
{2v^2(v^2-1)(1-u)}\ \epsilon,
\\ \nn &&K_{\varphi 1}\propto -\frac{v}{\sqrt{1-1/v^2-u}} - \frac{1+v^2(2u-1)}
{2(v^2-1)\sqrt{v^2(1-u)-1}(1-u)}\ \epsilon,
\\ \nn &&K_{\varphi 2}\propto 1-u +
\frac{\left(1-v^2(1-u)\right)u}{v^2-1}\ \epsilon,
\\ \nn &&K_{\varphi 3}\propto 1-v^2(1-u) + 2v^2u\left(1-\frac{v^2u}{v^2-1}\right)\epsilon
.\eea

\end{appendix}

\end{document}